\documentclass[%
 aip,
 sd,%
 amsmath,amssymb,
 reprint,%
]{revtex4-2}

\usepackage{multirow}
\usepackage{graphicx}
\usepackage{bm}
\usepackage{hyperref}

\usepackage{graphicx, textcomp, amssymb, mathrsfs}

\makeatletter
\newcommand\xleftrightarrow[2][]{%
\ext@arrow 9999{\longleftrightarrowfill@}{#1}{#2}}
\newcommand\longleftrightarrowfill@{%
\arrowfill@\leftarrow\relbar\rightarrow}
\makeatother

\usepackage{epstopdf}

\begin{document}

\title{Generalized Distribution Function of Relaxation Times with the Davidson-Cole Model as a Kernel}

\author{Anis Allagui$^*$}
\email{aallagui@sharjah.ac.ae}
\affiliation{Dept. of Sustainable and Renewable Energy Engineering, University of Sharjah, Sharjah, P.O. Box 27272, United Arab Emirates}
\altaffiliation[Also at ]{Center for Advanced Materials Research, Research Institute of Sciences and Engineering, University of Sharjah, Sharjah, P.O. Box 27272,  United Arab Emirates}
\affiliation{Dept. of Electrical and Computer Engineering, Florida International University, Miami, FL33174, United States}

\author{Ahmed Elwakil} 
\affiliation{
Dept. of Electrical Engineering, 
University of Sharjah, PO Box 27272, Sharjah, United Arab Emirates 
}

\begin{abstract}

In this paper we propose a generalized distribution function of relaxation times (DFRT) considering the   Davidson-Cole  model as an elementary process instead of the standard Debye model. The distribution function is retrieved  from the inverse of the  generalized Stieltjes transform expressed in terms of iterated Laplace transforms. We derive computable analytical expressions of the  generalized DFRT  for some of the most known normalized impedance (or admittance) models including the constant phase element, the Davidson-Cole,   Havriliak-Negami   and the Kohlrausch-Williams-Watts models.  
    


\end{abstract}

\maketitle

\section{Introduction}

When  a linear system is subjected to a frequency-domain input signal $X(s)\;(s=i\omega$, $\omega$ is the angular frequency) and resulting in a frequency-domain output signal $Y(s)$, the transfer function describing the dynamics of the system can be written as the ratio:
\begin{equation}
Z(s) = \frac{Y(s)}{X(s)}
\end{equation}
If $Z(s)$ corresponds to an electrical impedance, then $Y(s)$ is a voltage signal and $X(s)$ is a current signal. When used for treating capacitive systems with a   single time constant $\tau$, the function $Z(s)$ is given by:
\begin{equation}
Z_D(s) = R_{s}+ \frac{R_p}{1+s \tau}
\label{eq:Z}
\end{equation}
where $R_s$ and $R_p$ are two resistive constants (in units of Ohm). 
The fraction: 
\begin{equation}
Q_D(s) = ({1+s \tau})^{-1}
\end{equation}
is  the normalized or reduced Debye model, originally used for describing the frequency-dependent complex susceptibilities of dielectrics. In Nyquist representation of real vs. imaginary parts of impedance, Eq.\;\ref{eq:Z} shows a symmetric   semi-circular  shape, while its time response obtained by inverse Laplace transform (LT) corresponds to an exponential decay function:
\begin{equation}
z_D(t) = R_s\, \delta(t) + R_p\, \tau^{-1}{e^{-t/\tau}} 
\end{equation}
The convolution of the time-domain functions $z_D(t)$ with the current $i(t)$ gives the resulting voltage $v(t)$ as:
\begin{equation}
v(t) = (z_D \circledast i) (t) 
= \int\limits_{-\infty}^{+\infty} z_D(t-\xi) i(\xi)  d\xi 
\end{equation}

However, if the system under consideration is more complex than the idealized one, e.g. when the system is porous, amorphous or consists of disordered, inhomogeneous structures, the Debye model   is known to be  insufficient for capturing  accurately   its frequency behavior.  
  By applying the principle of superposition, and with the assumption that linearity still holds \cite{florsch2012direct}, the impedance can be expressed by the integral:
\begin{equation}
Z_D^*(s) = R_s + R_p\int\limits_0^{\infty} \frac{g(\tau)}{1+s \tau} d\tau
\label{eqZs}
\end{equation}
  where $g(\tau)$ is a distribution function of relaxation times (DFRT).
In normalized form, we rewrite Eq.\;\ref{eqZs} as:
\begin{equation}
Q_D^*(s) = \frac{Z_D^*(s) - R_s}{R_p} = \int\limits_0^{\infty} \frac{g(\tau)}{1+s \tau} d\tau
\label{eqZns}
\end{equation}
With this approach it is assumed that the relaxing macroscopic system consists of a large number of subsytems, each of which, upon the application of a step input, relaxes exponentially to zero   with its own single relaxation time constant \cite{ciucci2015analysis}. There have been several efforts to construct reliable and robust numerical algorithms for retrieving the DFRT of a system from its spectral response (Eq.\ref{eqZns}), including Fourier transform techniques \cite{schichlein2002deconvolution, boukamp2015fourier}, Tikhonov regularization \cite{gavrilyuk2017use, paul2021computation, shanbhag2020relaxation}, Max Entropy \cite{horlin1998deconvolution, horlin1993maximum}, genetic algorithm \cite{tesler2010analyzing}, etc. \cite{boukamp2020distribution, effendy2020analysis, liu2020deep, ciucci2015analysis}.

In general terms, Eq.\ref{eqZns} should be rather viewed   as:
\begin{equation}
Q_D^*(s) = \int\limits_0^{\infty}  {g(\tau)} k({s, \tau}) d\tau
\label{eqZnsGen}
\end{equation}
where $k({s, \tau})$ is the kernel of the integral. In the case of the elementary Debye response, we have $k({s, \tau}) = ({1+s\tau})^{-1}$.   
 But there are several possible extensions of the  Debye model that can be  considered. These include for instance the Cole-Cole (CC) model \cite{cole1941dispersion},
 \begin{equation}
Q_{\text{CC}}(s) = ({1+(s\tau_{\alpha})^{\alpha}})^{-1} \quad (0<\alpha \leqslant 1)
\label{eq:Qsalpha0}
\end{equation}
 the Davidson-Cole (DC) model \cite{davidson1951dielectric}, 
 \begin{equation}
Q_{\text{DC}}(s) = {(1+s\tau_{\beta})^{-\beta}} \quad (0< \beta \leqslant 1)
\label{eq:QsalphaCD}
\end{equation}
 and the Havriliak-Negami (HN) model \cite{havriliak1966complex},   
 \begin{equation}
Q_{\text{HN}}(s) = {(1+(s\tau_{{\nu}})^{\alpha})^{-\beta}} \quad (0< \alpha, \beta \leqslant 1) 
\label{eq:QsalphaHN}
\end{equation} 
which are 
capable of describing  a wide range of depressed and asymmetric  semi-circular shapes in the complex plane with different levels of accuracy.
In the time domain the relaxation processes corresponding to these models are non-exponential. By inverse LT \cite{prabhakar1971singular}, the response  of the CC function to a step excitation is \cite{garrappa2016models}: 
\begin{equation}
z_{\text{CC}}(t) = \tau_{\alpha}^{-1}({t}/{\tau_{\alpha}})^{\alpha-1} E_{\alpha,\alpha}^1\left[ -(t/\tau_{\alpha})^{\alpha} \right]
\label{eq41}
\end{equation}
where \cite{prabhakar1971singular, saxena2004generalized}
\begin{equation}
{E}_{\alpha,\beta}^{\gamma} ( z ) := \sum\limits_{k=0}^{\infty} \frac{(\gamma)_k}{\Gamma(\alpha k + \beta)} \frac{z^k}{k!} \quad (\alpha,\beta, \gamma \in \mathbb{C}, \mathrm{Re}({\alpha})>0)
\label{eqML}
\end{equation}
(with $(\gamma)_k = \gamma(\gamma+1)\ldots(\gamma+k-1) =\Gamma(\gamma+k)/\Gamma(\gamma)$) is the three-parameter Mittag-Leffler (ML) function. The   response   functions of the DC and HN models are \cite{garrappa2016models}: 
\begin{align}
z_{\text{DC}}(t)
 &= \tau_{\beta}^{-1}({t}/{\tau_{\beta}})^{\beta-1} E_{1,\beta}^{\beta}\left[ -t/\tau_{\beta} \right] \\
& =
\frac{1}{\Gamma(\beta)}\tau_{\beta}^{-1}({t}/{\tau_{\beta}})^{\beta-1}
e^{-t/\tau_{\beta} }
\label{eq:Agammat0}
\end{align}
and 
\begin{equation}
z_{\text{HN}}(t)= \tau_{\nu}^{-1}({t}/{\tau_{\nu}})^{\alpha \beta-1} E_{\alpha,\alpha \beta}^{\beta}\left[ -(t/\tau_{\nu})^{\alpha} \right]
\label{eqHH}
\end{equation}
respectively.

Returning to  Eq.\ref{eqZns}, Florsch, Revil, and Camerlynck \cite{florsch2014inversion} proposed for instance the possibility of using the HN model as the kernel of the integral. This gives  for the normalized impedance function:
\begin{equation}
Q^*_{\text{HN}}(s) = \int\limits_{0}^{\infty} \frac{g(\tau)}{(1+ (s \tau_{\nu})^{\alpha})^{\beta}} d\tau \quad ( 0< \alpha, \gamma  \leqslant 1)
\label{eq:ZfHN}
\end{equation}
The problem here is  to retrieve $g(\tau)$ assuming that the parameters $\alpha$ and $\beta$ are known.  The goal   is that one may want to decompose the spectral response of the system using for example the Warburg model as a basis function. This can be obtained by setting $\beta=1$ and $\alpha=0.5$ in Eq.\;\ref{eq:ZfHN}. In any case the choice of   values for the parameters $\alpha$ and $\beta$ should be justified in advance based on knowledge of the underlying physics. 
The authors carried out their analysis for recovering the distribution function $g(\tau)$ from Eq.\;\ref{eq:ZfHN} by numerical inversion algorithms given the ill-posedness of the problem.

In the same line of   thought, the purpose of this work is to present a general procedure for obtaining  \emph{analytically} the DFRT associated with an impedance model when the elementary relaxation process is taken to be the DC model (Eq.\ref{eq:Zf02}) instead of the Debye model.  The motivation for this generalization applies for instance to macroscopic systems of self-similar character  at a microscopic level, with the DC type of response in the frequency  domain \cite{nigmatullin1997cole, ieee}.

\section{Theory}

Consider the integral transform:
 \begin{equation}
Q^*_{\text{DC}}(s) = \int\limits_{0}^{\infty} \frac{g(\tau)}{(1+ s \tau)^{p}} d\tau
\label{eq:Zf02}
\end{equation} 
which is equivalent to Eq.\;\ref{eqZnsGen} with the kernel $k({s, \tau}) = ({1+s\tau})^{-p}$. 
We will drop both the subscript $"\text{DC}"$ and the superscript $"^{*}"$ from now on for ease of notation.  
Using the change of variable $\lambda=\tau^{-1}$, and the distribution function:
\begin{equation}
D(\lambda) = \lambda^{p-2} g(\lambda^{-1})
\label{eq:Dlambda}
\end{equation}
 Eq.\;\ref{eq:Zf02} is rewritten as
\begin{equation}
Q(s) = \int\limits_{0}^{\infty} \frac{D (\lambda) }{( s + \lambda )^{p}} d\lambda
\label{eq:LTLT}
\end{equation}
which is the generalized Stieltjes transform (ST) \cite{bateman1954tables}. With $p=1$ we recover the standard ST \cite{drt}. 
Following the procedure of Schwarz \cite{schwarz2005generalized}, 
by using the Mellin transform of a function $f(x)$ defined as $\mathcal{M}[f(x);s] = \int_0^{\infty} x^{s-1} f(x) dx\;(x>0)$, we rewrite the kernel ${( s + \lambda )^{-p}}$ in Eq.\;\ref{eq:LTLT} as:
\begin{equation}
{( s + \lambda )^{-p}}  = \frac{1}{\Gamma(p)} \int\limits_0^{\infty} x^{p-1} e^{-x( s+\lambda )} dx
\end{equation}
Plugging it in Eq.\;\ref{eq:LTLT} makes $Q (s)$   to be:
 \begin{equation}
Q(s) = \frac{1}{\Gamma(p)} \int\limits_0^{\infty} x^{p-1} e^{-xs} \int\limits_0^{\infty} D (\lambda) e^{-x \lambda} d\lambda \, dx
 \end{equation}
From the definition of the LT, we recognize that:
\begin{equation}
Q (s) = \frac{1}{\Gamma(p) } 
\mathcal{L} \left[ x^{p-1} \mathcal{L} [ D(\lambda);x] ;s \right]
 \end{equation}
 and with the inverse LT, we recover $D(\lambda)$ from the iterations:
\begin{equation}
D(\lambda) = \Gamma(p) \mathcal{L}^{-1} \left[ 
x^{1-p} \mathcal{L}^{-1} [Q(s);x]; \lambda
\right]
\label{eq:DlambdaLL}
\end{equation}
The DFRT $g(\tau)$ is then obtained from Eq.\;\ref{eq:Dlambda} as:
\begin{equation}
g(\tau) = \tau^{p-2} D(\tau^{-1})
\label{Eq:g}
\end{equation} 
We verify that for $p=1$, $D(\lambda) = \mathcal{L}^{-1} \left[  \mathcal{L}^{-1} [Q(s);x]; \lambda\right]$ and $g(\tau) = \tau^{-1} D(\tau^{-1})$ as they should be \cite{macdonald1956linear, drt}.

\section{Examples}
\label{sec:examples}

\subsection{Constant phase element}
\begin{figure}[t]
\begin{center}
\includegraphics[width=0.985\columnwidth]{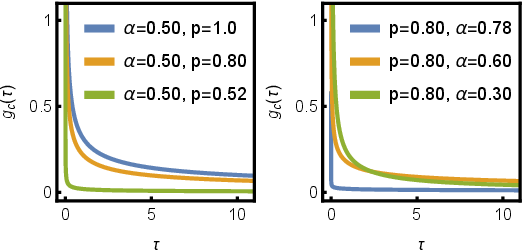}
\caption{Plots of $g_c(\tau)$ in Eq.\;\ref{eq:gctau} for different combinations of the parameters $\alpha$ and $p$, and with $\tau_c=1$}
\label{fig1}
\end{center}
\end{figure}

As a first example we consider the  case of the reduced impedance of the constant phase element given by:
\begin{equation}
Q_{c}(s) =  {(s\tau_{c})^{-\alpha}}, \quad 0<\alpha \leqslant 1
\label{eq:CPE} 
\end{equation}
where $\tau_{c}$ is a  positive constant. This is basically the case of the CC model in Eq.\;\ref{eq:Qsalpha0} when $s\tau_{\alpha} \gg 1$. Using Eq.\;\ref{eq:DlambdaLL} followed by Eq.\;\ref{Eq:g} we obtain:
\begin{equation}
g_c(\tau) = \frac{\Gamma(p)}{\Gamma(\alpha) \Gamma(p- \alpha)} \tau_c^{-\alpha} \tau^{\alpha-1}, \;\; p>\alpha
\label{eq:gctau}
\end{equation}
Given that for physical reasons the distribution function $g(\tau)$ cannot be negative, we need to ensure the condition $0<\alpha<p<1$. Otherwise, the term $  {\Gamma(p)}/[{\Gamma(\alpha) \Gamma(p- \alpha)}]$ should be positive. 
We note that for  $p=1$ (i.e. Debye model as a kernel in Eq.\;\ref{eqZnsGen}: $k({s, \tau}) = ({1+s\tau})^{-1}$), $g_c(\tau)$ simplifies to
\begin{equation}
g_{c}(\tau) = \pi^{-1} \sin(\alpha\pi) \tau_{c}^{-\alpha} \tau^{\alpha-1} 
\end{equation}
 where we used  Euler's reflection formula for the gamma function,   $  {\pi}/{\sin(\pi z)} = \Gamma(z) \Gamma(1-z) $ ($z \notin \mathbb{Z}$). 
This means that a CPE impedance of parameters $\tau_c$ and $\alpha$ can   equally be constructed from the integral:
\begin{equation}
Q_c(s) ={(s\tau_{c})^{-\alpha}} = \frac{\tau_{c}^{-\alpha}}{\Gamma(\alpha) \Gamma(1- \alpha)}  \int\limits_0^{\infty} \frac{\tau^{\alpha-1}}{1+s \tau} d\tau 
\end{equation}
using the Debye model as a basis function   \cite{drt}, or from the integral:
\begin{equation}
Q_c(s) ={(s\tau_{c})^{-\alpha}}= \frac{\Gamma(p)\, \tau_c^{-\alpha}}{\Gamma(\alpha) \Gamma(p- \alpha)}  \int\limits_0^{\infty} \frac{\tau^{\alpha-1}}{(1+s \tau)^{p}} d\tau 
\end{equation}
using the DC model (of any parameter $p>\alpha$) as a basis function. 

In Fig.\;\ref{fig1} we show plots of $g_c(\tau)$ given by Eq.\;\ref{eq:gctau} for different combinations of the parameters $\alpha$ and $p$, and with $\tau_c=1$. We can see that as the value of $p$ tends to that of $\alpha$, the distribution function becomes narrower and less disperse, which can be interpreted  as a more efficient way for reconstructing the CPE impedance.

\subsection{Davidson-Cole model}

\begin{figure}[b]
\begin{center}
\includegraphics[width=0.985\columnwidth]{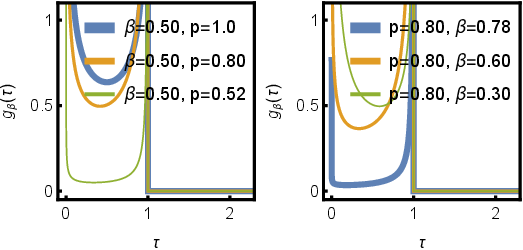}
\caption{Plots of $g_{\beta}(\tau)$ in Eq.\;\ref{eq:gbeta} for different combinations of the parameters $\beta$ and $p$, and with $\tau_{\beta}=1$}
\label{fig2}
\end{center}
\end{figure}

If we consider the impedance function itself to be of a DC type, i.e.
\begin{equation}
Q_{\beta}(s) = {(1+s\tau_{\beta})^{-\beta}}    =\int\limits_{0}^{\infty} \frac{g_{\beta}(\tau)}{(1+ s \tau)^{p}} d\tau
\label{eq:31}
\end{equation}
with $0< \beta,p \leqslant 1$ and $\tau_{\beta}$ a reference   time constant, we find the    $g_{\beta}(\tau)$ function to be
\begin{align}
g_{\beta}(\tau)&= \frac{\Gamma (p ) \tau_{\beta}^{-\beta} }{\Gamma (\beta ) \Gamma (p -\beta )}
 \tau ^{p -2}  
\left({\tau^{-1} }-{\tau_{\beta}^{-1}}\right)^{p-\beta -1}, 
\text{ for } \tau< \tau_{\beta} \nonumber \\
& = 0,   \text{ for } \tau \geqslant \tau_{\beta}
\label{eq:gbeta}
\end{align} 
We verify that for $p=1$,  $g_{\beta}(\tau)$ simplifies to \cite{drt}: 
\begin{align}
g_{\beta}(\tau) &=\frac{\sin (\beta \pi   )  }{\pi   } \tau_{\beta}^{-\beta}  \tau^{-1} (\tau^{-1} -\tau_{\beta}^{-1})^{-\beta }, \text{ for } \tau< \tau_{\beta} \nonumber \\
& = 0,  \text{ for } \tau \geqslant \tau_{\beta}
\end{align}
and for $p=\beta$, we obtain evidently the Dirac delta function:
\begin{equation}
g_{\beta}(\tau) =  \delta(\tau-\tau_{\beta})
\end{equation}

In Fig.\;\ref{fig2} we show the highly asymmetric  plots of $g_{\beta}(\tau)$ in Eq.\;\ref{eq:gbeta} for different combinations of the parameters $\beta$ and $p$, and with $\tau_{\beta}=1$. We have the same remark as for the case of the CPE: a DC-type of  impedance can be reconstructed from the continuous linear superposition of Debye models, but also from other DC models. Furthermore, Eq.\;\ref{eq:31}, with $g_{\beta}(\tau)$ given by Eq.\;\ref{eq:gbeta}, shows one possible mapping relationship  between two  DC models with different parameters.

\begin{figure*}[t]
\begin{center}
\includegraphics[width=0.23\textwidth,angle=-90]{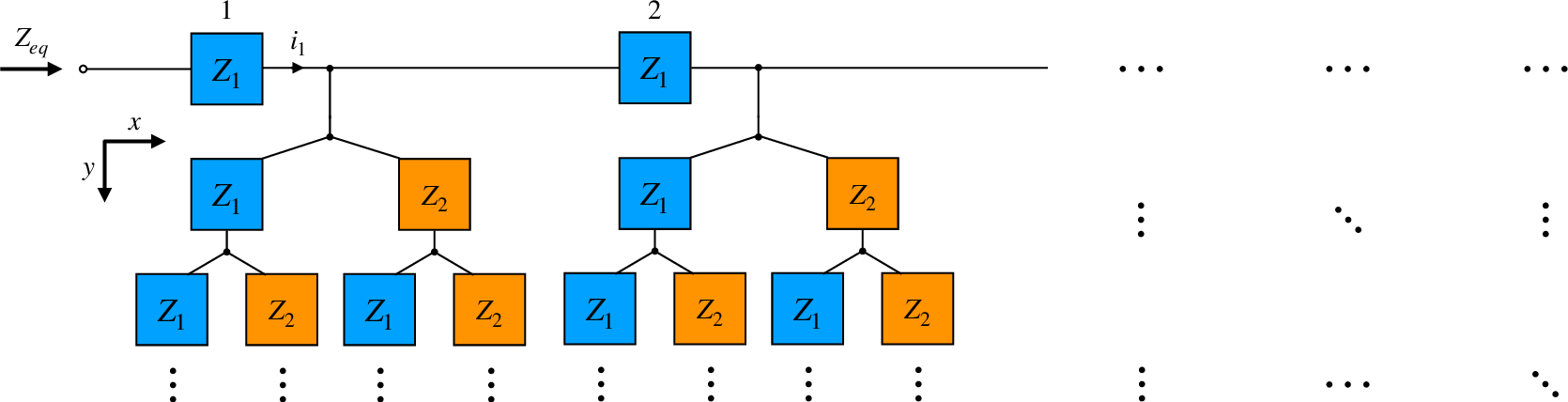}
\caption{Schematic of a 1D semi-infinite two-impedance  self-similar ladder network in the $x$-direction, coupled with a semi-infinite bifurcating binary tree network in the $y$-direction.}
\label{fig3}
\end{center}
\end{figure*}
 
\subsection{Havriliak–Negami model}

Now we show the inversion procedure for the the case of the HN model, i.e.: 
\begin{equation}
Q_{\nu}(s) = {(1+ (s\tau_{\nu})^{\alpha})^{-\beta}}    =\int\limits_{0}^{\infty} \frac{g_{\nu}(\tau)}{(1+ s \tau)^{p}} d\tau
\label{eq:HNintegral}
\end{equation}
from which the CC model can be a recovered as a special case by setting $\beta=1$.   
For this example, it is more convenient to use the representation of $Q_{\nu}(s)$ in terms of the Fox $H$-function, as we have shown   in more detail  in a separate contribution of ours \cite{drt} for the case of $k({s, \tau}) = ({1+s\tau})^{-1}$. 
We recall that the $H$-function  of order 
$(m,n,p,q)\in \mathbb{N}^4$, ($0 \leqslant n \leqslant p$, $1 \leqslant m \leqslant q$) 
and with parameters 
  $A_j \in \mathbb{R}_+ \;(j=1,\ldots,p)$, $B_j \in \mathbb{R}_+\; (j=1,\ldots,q) $, 
$a_j \in\mathbb{C}\;(j=1,\ldots,p)$ and $ b_j \in \mathbb{C}\; (j=1,\ldots,q)$ 
is defined for $z \in \mathbb{C},\;z\neq 0$ by the contour  integral  \cite{fox1961g, mathai2009h, kilbas2004h, mathai1978h}:
\begin{equation}
H^{m,n}_{p,q}\left[ z|^{(a_1,A_1),\ldots,(a_p,A_p)}_{(b_1,B_1),\ldots, (b_q,B_q)} \right] =\frac{1}{2\pi i} \int_L h(s) z^{-s} ds
\label{eq:H}
\end{equation}
 where the integrand $h(s)$ is given by:
 \begin{equation}
h(s) = \frac{\left\{\prod\limits_{j=1}^m \Gamma(b_j + B_j s)\right\}  \left\{\prod\limits_{j=1}^n \Gamma(1-a_j - A_j s)\right\}}
{\left\{\prod\limits_{j={m+1}}^q \Gamma(1-b_j - B_j s)\right\} \left\{\prod\limits_{j={n+1}}^p \Gamma(a_j + A_j s)\right\}}
\end{equation}
In Eq.\;\ref{eq:H}, $z^{-s}=\exp \left[ -s (\ln|z|+ i \arg z) \right] $ and $\arg z$ is not necessarily the principal value. 
 The contour of integration $L$ is a suitable contour separating the poles  
 of  $\Gamma(b_j+ B_j s)$ ($j=1,\ldots,m$) from the poles 
 of   $\Gamma (1-a_{j} - A_{j} s)$ ($j=1,\ldots,n$).
 An empty product is always interpreted as unity. 
 We rewrite $Q_{\nu}(s)$ as
\begin{equation}
Q_{\nu}(s) = \frac{1}{\Gamma(\beta)} 
H^{1,1}_{1,1}\left[ (s\tau_{\nu})^{\alpha}  \left|
\begin{array}{c}
(1-\beta,1)    \\
(0,1) \hfill   \\
\end{array}
\right.\right]
\label{eq:QsalphaHN}
\end{equation} 
using the formula\;\cite{mathai2009h}:
\begin{align}
(1- z)^{-\alpha} =  \frac{1}{\Gamma(\alpha)} 
H^{1,1}_{1,1}\left[ - z  \left|
\begin{array}{c}
(1-\alpha,1)    \\
(0,1) \hfill   \\
\end{array}
\right.\right]
\end{align}  
Knowing that the LT of an $H$-function is again an $H$-function, and is given by \cite{hilfer, glockle1993fox, mathai2009h}: 
\begin{align}
\label{eq:LTH}
\mathcal{L} 
&\left[ 
t^{\rho-1} H_{p,q}^{m,n}\left[ a t^{\sigma} \left|
\begin{array}{c}
(a_p,A_p)  \\
(b_q,B_q)  \\
\end{array}
\right.\right];  u
\right] \\ \nonumber
 &= 
  u^{-\rho} 
H^{m,n+1}_{p+1,q}\left[ a  u^{-\sigma}\left|
\begin{array}{c}
(1-\rho,\sigma), (a_1,A_1), \ldots, (a_p,A_p)   \\
(b_1,B_1), \ldots, (b_q,B_q)  \\
\end{array}
\right.\right]
\end{align}
 and the inverse LT  is given by \cite{mathai2009h}:
\begin{align}
\mathcal{L}^{-1} 
&\left[ 
  u^{-\rho} 
H_{p,q}^{m,n}\left[ a  u^{\sigma}\left|
\begin{array}{c}
(a_p,A_p)  \\
(b_q,B_q)  \\
\end{array}
\right.\right]; t
\right] \label{eq:ILT} \\ \nonumber
 &= t^{\rho-1} 
H_{p+1,q}^{m,n}\left[ a t^{-\sigma}\left|
\begin{array}{c}
(a_p,A_p), \ldots, (a_1,A_1), (\rho,\sigma)  \\
(b_1,B_1),\ldots,(b_q,B_q)  \\
\end{array}
\right.\right]
\end{align}
we find, after employing some identities between $H$-functions that \cite{kilbas2004h,mathai2009h}: 
\begin{equation}
g_{\nu}(\tau) = \frac{\Gamma(p)}{\tau_{\nu}\, \Gamma(\beta)} H^{1,1}_{2,2}\left[ (\tau/\tau_{\nu})^{\alpha} \left|
\begin{array}{c}
(1-1/\alpha,1), (p-1,\alpha)  \\
(\beta-1/\alpha,1), (0,\alpha)  \hfill \\
\end{array}
\right.\right]
\label{eq:gHN}
\end{equation}
This general expression includes as special cases  the DFRT for the DC and CC reduced impedances for different values of $p$ ($0< p \leqslant 1$).

\begin{figure}[t]
\begin{center}
\includegraphics[width=0.985\columnwidth]{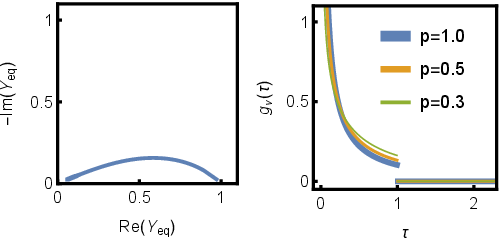}
\caption{In (a) Nyquist plot of real vs. imaginary parts of $Y_{eq} = 1/\sqrt{1+\sqrt{s\tau_{eq}}}$ for $\tau_{eq}=1$, frequency from 1\,mHz to 10\,kHz; In (b) plots of $g_{\nu}(\tau)$ in Eq.\;\ref{eq:gHN} for $\alpha=\beta=0.5$, $\tau_{\nu}=1$, and  different values of $p$ (1.0, 0.5 and 0.3)}
\label{fig4}
\end{center}
\end{figure}

As a particular example, 
let us consider the electrical circuit model shown in Fig.\;\ref{fig3}, which represents a 1D semi-infinite two-impedance ($Z_1$ and $Z_2$) self-similar ladder network in the $x$-direction, coupled with a semi-infinite bifurcating binary tree network in the $y$-direction.  Its equivalent impedance function as viewed from the input terminal of the network is given by (ladder network) \cite{ieee}:
\begin{equation}
2 Z_{eq}^L=Z_1 \left( 1+\sqrt{1+4Z_{eq}^{T}/Z_1} \right)
\end{equation}
where $Z_{eq}^T=\sqrt{Z_1 Z_2}$ (binary tree). This leads to:
\begin{equation}
2 Z_{eq}^L=Z_1 \left( 1+\sqrt{1+4 \sqrt{Z_2/Z_1}} \right)
\end{equation}
If we take $Z_1$ to be a resistance $R$ and $Z_2$ to be an inductance of impedance $sL$, then we obtain a normalized impedance $Z^{L*}_{eq}$ as:
\begin{equation}
Z^{L*}_{eq} = \frac{2 Z_{eq}^L}{R}-1 = \sqrt{1+\sqrt{s\tau_{eq}}} 
\end{equation}
where $\tau_{eq}=16{L/R}$. The reciprocal  gives a normalized admittance $Y^{L*}_{eq} = 1/Z^{L*}_{eq}$ of the HN type where $\alpha=0.5$ and $\beta=0.5$ (Eq.\;\ref{eq:HNintegral}). In Fig.\;\ref{fig4}(a), we show the plot of $Y^{L*}_{eq}$ ($Y_{eq}$ in the figure) in Nyquist format, with $\tau_{eq}=1$ over the frequency range 1\,mHz to 10\,kHz. 
Plots of $g_{\nu}(\tau)$ (Eq.\;\ref{eq:gHN}) with $\alpha=0.5$, $\beta=0.5$, $\tau_{\nu}=1$, and three different values of $p$ (1.0, 0.5 and 0.3) are given in Fig.\;\ref{fig4}(b).

\subsection{Kohlrausch-Williams-Watts model}

We conclude with the Kohlrausch-Williams-Watts (KWW) model as a fourth example of our study \cite{williams1970non}. The KWW model is actually introduced starting from its time-domain relaxation function, which is given in normalized form by \cite{garrappa2016models}:
\begin{equation}
\rho_{\kappa}(t) = e^{-(t/\tau_{\kappa})^{\kappa}},\;\; t\geqslant 0,\; 0<\kappa \leqslant 1
\label{eq:stretchExp}
\end{equation}
This function is also known as the stretched exponential function with an exponent $\kappa$ and time constant $\tau_{\kappa}$. While the response function $z(t)$ and the normalized impedance function  $Q(s)$ are a LT pair,  $\rho_{\kappa}(t)$  and $Q(s)$ are related via \cite{garrappa2016models}:
\begin{equation}
\rho_{\kappa}(t) = 1- \mathcal{L}^{-1}[ Q_{\kappa}(s)/s; t] = \mathcal{L}^{-1}[ (1-Q_{\kappa}(s))/s; t] 
\label{eq:fott}
\end{equation}
This comes from $z(t) = -d \rho(t)/dt$. 
With the $H$-function identity: 
\begin{equation}
e^{-(t/\tau_{\kappa})^{\kappa}} = 
H^{1,0}_{0,1}\left[ (t/\tau_{\kappa})^{\kappa}  \left|
\begin{array}{c}
- \hfill    \\
(0,1) \hfill   \\
\end{array}
\right.\right]
\end{equation}
one can obtain by applying the LT (Eq.\;\ref{eq:LTH}) to both sides of Eq.\;\ref{eq:fott} \cite{hilfer}:
\begin{equation}
Q_{\kappa}(s)= 1-
H^{1,1}_{1,1}\left[ (s \tau_{\kappa})^{\kappa}  \left|
\begin{array}{c}
(1,1) \hfill    \\
(1,\kappa) \hfill   \\
\end{array}
\right.\right]
\end{equation}
From Garrappa et al. \cite{garrappa2016models}, it can be shown that the LT of the relaxation function being $(1-Q_{\kappa}(s))/s$ is in fact the ST of its corresponding spectral function. However, in order for us to apply directly Eq.\;\ref{eq:DlambdaLL}, we derive from Eq.\;\ref{eq:stretchExp} the response function of the system as:
\begin{equation}
z_{\kappa}(t) = - \frac{d\rho_{\kappa}(t)}{dt} =  
(\kappa / \tau_{\kappa})
(t/\tau_{\kappa})^{\kappa-1}
 e^{-(t/\tau_{\kappa})^{\kappa}}
\end{equation}
Applying the LT we verify that $Q_{\kappa}(s)$ is also equal to:
\begin{equation}
Q_{\kappa}(s)= 
\kappa \, 
H^{1,1}_{1,1}\left[ (s \tau_{\kappa})^{\kappa}  \left|
\begin{array}{c}
(0,1) \hfill    \\
(0,\kappa) \hfill   \\
\end{array}
\right.\right]
\label{eq:qofskappa}
\end{equation}
We then find from Eq.\;\ref{eq:DlambdaLL} and Eq.\;\ref{Eq:g}: 
\begin{align}
g_{\kappa}(\tau) &=  {\kappa\, \Gamma(p)}{\tau^{-1}  } H^{1,1}_{2,2}\left[ (\tau/\tau_{\kappa})^{\kappa} \left|
\begin{array}{c}
(1,\kappa), (p,\kappa)  \\
( 1,1), (1,\kappa)  \hfill \\
\end{array}
\right.\right] \\
&= {\kappa\, \Gamma(p)}{\tau^{-1}  } 
H^{1,0}_{1,1}
\left[ (\tau/\tau_{\kappa})^{\kappa} \left|
\begin{array}{c}
 (p,\kappa)  \\
( 1,1)  \hfill \\
\end{array}
\right.\right]
\label{eq:gkappa}
\end{align}

\begin{figure}[t]
\begin{center}
\includegraphics[width=0.985\columnwidth]{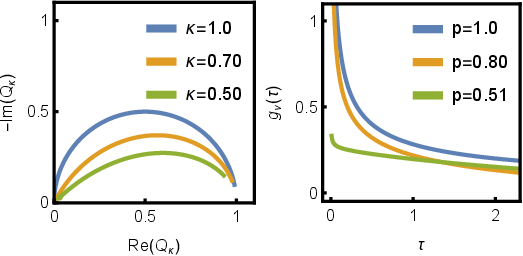}
\caption{In (a) Nyquist plot of real vs. imaginary parts of $Q_{\kappa}(s)$ (Eq.\;\ref{eq:qofskappa}) for $\tau_{\kappa}=1$ and different values of $\kappa$ (1.0, 0.7 and 0.5); the frequency range is 16\,mHz to 100\,Hz. 
 In (b) plots of $g_{\kappa}(\tau)$ in Eq.\;\ref{eq:gkappa} for $\kappa=0.5$, $\tau_{\nu}=1$, and  different values of $p$ (1.0, 0.8 and 0.51)}
\label{fig5}
\end{center}
\end{figure}

In Fig.\;\ref{fig5}(a) we show the Nyquist plots of Eq.\;\ref{eq:qofskappa}  for $\tau_{\kappa}=1$ and for different values of $\kappa$ (1.0, 0.7 and 0.5). In Fig.\;\ref{fig5}(b), we show plots of $g_{\kappa}(\tau)$ in Eq.\;\ref{eq:gkappa} for the simple case of $\kappa=0.5$, $\tau_{\kappa}=1$, and for different values of $p$ (1.0, 0.8 and 0.51). 

\section{Conclusion}

A generalized distribution function of relaxation times based on the elementary Davidson-Cole model for treating impedance functions is presented. The inversion procedure involves the generalized Stieltjes transform, which can be executed relatively easily using the Fox's $H$-function and its properties. We tested and validated the procedure for a few cases of well-known   functions.

\section*{References}


\begin{thebibliography}{33}%
\makeatletter
\providecommand \@ifxundefined [1]{%
 \@ifx{#1\undefined}
}%
\providecommand \@ifnum [1]{%
 \ifnum #1\expandafter \@firstoftwo
 \else \expandafter \@secondoftwo
 \fi
}%
\providecommand \@ifx [1]{%
 \ifx #1\expandafter \@firstoftwo
 \else \expandafter \@secondoftwo
 \fi
}%
\providecommand \natexlab [1]{#1}%
\providecommand \enquote  [1]{``#1''}%
\providecommand \bibnamefont  [1]{#1}%
\providecommand \bibfnamefont [1]{#1}%
\providecommand \citenamefont [1]{#1}%
\providecommand \href@noop [0]{\@secondoftwo}%
\providecommand \href [0]{\begingroup \@sanitize@url \@href}%
\providecommand \@href[1]{\@@startlink{#1}\@@href}%
\providecommand \@@href[1]{\endgroup#1\@@endlink}%
\providecommand \@sanitize@url [0]{\catcode `\\12\catcode `\$12\catcode
  `\&12\catcode `\#12\catcode `\^12\catcode `\_12\catcode `\%12\relax}%
\providecommand \@@startlink[1]{}%
\providecommand \@@endlink[0]{}%
\providecommand \url  [0]{\begingroup\@sanitize@url \@url }%
\providecommand \@url [1]{\endgroup\@href {#1}{\urlprefix }}%
\providecommand \urlprefix  [0]{URL }%
\providecommand \Eprint [0]{\href }%
\providecommand \doibase [0]{https://doi.org/}%
\providecommand \selectlanguage [0]{\@gobble}%
\providecommand \bibinfo  [0]{\@secondoftwo}%
\providecommand \bibfield  [0]{\@secondoftwo}%
\providecommand \translation [1]{[#1]}%
\providecommand \BibitemOpen [0]{}%
\providecommand \bibitemStop [0]{}%
\providecommand \bibitemNoStop [0]{.\EOS\space}%
\providecommand \EOS [0]{\spacefactor3000\relax}%
\providecommand \BibitemShut  [1]{\csname bibitem#1\endcsname}%
\let\auto@bib@innerbib\@empty
\bibitem [{\citenamefont {Florsch}, \citenamefont {Camerlynck},\ and\
  \citenamefont {Revil}(2012)}]{florsch2012direct}%
  \BibitemOpen
  \bibfield  {author} {\bibinfo {author} {\bibfnamefont {N.}~\bibnamefont
  {Florsch}}, \bibinfo {author} {\bibfnamefont {C.}~\bibnamefont
  {Camerlynck}},\ and\ \bibinfo {author} {\bibfnamefont {A.}~\bibnamefont
  {Revil}},\ }\bibfield  {title} {\enquote {\bibinfo {title} {Direct estimation
  of the distribution of relaxation times from induced-polarization spectra
  using a fourier transform analysis},}\ }\href@noop {} {\bibfield  {journal}
  {\bibinfo  {journal} {Near Surf. Geophys.}\ }\textbf {\bibinfo {volume}
  {10}},\ \bibinfo {pages} {517--531} (\bibinfo {year} {2012})}\BibitemShut
  {NoStop}%
\bibitem [{\citenamefont {Ciucci}\ and\ \citenamefont
  {Chen}(2015)}]{ciucci2015analysis}%
  \BibitemOpen
  \bibfield  {author} {\bibinfo {author} {\bibfnamefont {F.}~\bibnamefont
  {Ciucci}}\ and\ \bibinfo {author} {\bibfnamefont {C.}~\bibnamefont {Chen}},\
  }\bibfield  {title} {\enquote {\bibinfo {title} {Analysis of electrochemical
  impedance spectroscopy data using the distribution of relaxation times: A
  bayesian and hierarchical bayesian approach},}\ }\href@noop {} {\bibfield
  {journal} {\bibinfo  {journal} {Electrochim. Acta}\ }\textbf {\bibinfo
  {volume} {167}},\ \bibinfo {pages} {439--454} (\bibinfo {year}
  {2015})}\BibitemShut {NoStop}%
\bibitem [{\citenamefont {Schichlein}\ \emph {et~al.}(2002)\citenamefont
  {Schichlein}, \citenamefont {M{\"u}ller}, \citenamefont {Voigts},
  \citenamefont {Kr{\"u}gel},\ and\ \citenamefont
  {Ivers-Tiff{\'e}e}}]{schichlein2002deconvolution}%
  \BibitemOpen
  \bibfield  {author} {\bibinfo {author} {\bibfnamefont {H.}~\bibnamefont
  {Schichlein}}, \bibinfo {author} {\bibfnamefont {A.~C.}\ \bibnamefont
  {M{\"u}ller}}, \bibinfo {author} {\bibfnamefont {M.}~\bibnamefont {Voigts}},
  \bibinfo {author} {\bibfnamefont {A.}~\bibnamefont {Kr{\"u}gel}},\ and\
  \bibinfo {author} {\bibfnamefont {E.}~\bibnamefont {Ivers-Tiff{\'e}e}},\
  }\bibfield  {title} {\enquote {\bibinfo {title} {Deconvolution of
  electrochemical impedance spectra for the identification of electrode
  reaction mechanisms in solid oxide fuel cells},}\ }\href@noop {} {\bibfield
  {journal} {\bibinfo  {journal} {J. Appl. Electrochem.}\ }\textbf {\bibinfo
  {volume} {32}},\ \bibinfo {pages} {875--882} (\bibinfo {year}
  {2002})}\BibitemShut {NoStop}%
\bibitem [{\citenamefont {Boukamp}(2015)}]{boukamp2015fourier}%
  \BibitemOpen
  \bibfield  {author} {\bibinfo {author} {\bibfnamefont {B.~A.}\ \bibnamefont
  {Boukamp}},\ }\bibfield  {title} {\enquote {\bibinfo {title} {Fourier
  transform distribution function of relaxation times; application and
  limitations},}\ }\href@noop {} {\bibfield  {journal} {\bibinfo  {journal}
  {Electrochim. Acta}\ }\textbf {\bibinfo {volume} {154}},\ \bibinfo {pages}
  {35--46} (\bibinfo {year} {2015})}\BibitemShut {NoStop}%
\bibitem [{\citenamefont {Gavrilyuk}, \citenamefont {Osinkin},\ and\
  \citenamefont {Bronin}(2017)}]{gavrilyuk2017use}%
  \BibitemOpen
  \bibfield  {author} {\bibinfo {author} {\bibfnamefont {A.}~\bibnamefont
  {Gavrilyuk}}, \bibinfo {author} {\bibfnamefont {D.}~\bibnamefont {Osinkin}},\
  and\ \bibinfo {author} {\bibfnamefont {D.}~\bibnamefont {Bronin}},\
  }\bibfield  {title} {\enquote {\bibinfo {title} {The use of tikhonov
  regularization method for calculating the distribution function of relaxation
  times in impedance spectroscopy},}\ }\href@noop {} {\bibfield  {journal}
  {\bibinfo  {journal} {Russ. J. Electrochem.}\ }\textbf {\bibinfo {volume}
  {53}},\ \bibinfo {pages} {575--588} (\bibinfo {year} {2017})}\BibitemShut
  {NoStop}%
\bibitem [{\citenamefont {Paul}\ \emph {et~al.}(2021)\citenamefont {Paul},
  \citenamefont {Chi}, \citenamefont {Wu},\ and\ \citenamefont
  {Wu}}]{paul2021computation}%
  \BibitemOpen
  \bibfield  {author} {\bibinfo {author} {\bibfnamefont {T.}~\bibnamefont
  {Paul}}, \bibinfo {author} {\bibfnamefont {P.}~\bibnamefont {Chi}}, \bibinfo
  {author} {\bibfnamefont {P.~M.}\ \bibnamefont {Wu}},\ and\ \bibinfo {author}
  {\bibfnamefont {M.}~\bibnamefont {Wu}},\ }\bibfield  {title} {\enquote
  {\bibinfo {title} {Computation of distribution of relaxation times by
  tikhonov regularization for li ion batteries: usage of l-curve method},}\
  }\href@noop {} {\bibfield  {journal} {\bibinfo  {journal} {Sci. Rep.}\
  }\textbf {\bibinfo {volume} {11}},\ \bibinfo {pages} {12624} (\bibinfo {year}
  {2021})}\BibitemShut {NoStop}%
\bibitem [{\citenamefont {Shanbhag}(2020)}]{shanbhag2020relaxation}%
  \BibitemOpen
  \bibfield  {author} {\bibinfo {author} {\bibfnamefont {S.}~\bibnamefont
  {Shanbhag}},\ }\bibfield  {title} {\enquote {\bibinfo {title} {Relaxation
  spectra using nonlinear tikhonov regularization with a bayesian criterion},}\
  }\href@noop {} {\bibfield  {journal} {\bibinfo  {journal} {Rheol. Acta}\
  }\textbf {\bibinfo {volume} {59}},\ \bibinfo {pages} {509--520} (\bibinfo
  {year} {2020})}\BibitemShut {NoStop}%
\bibitem [{\citenamefont {H{\"o}rlin}(1998)}]{horlin1998deconvolution}%
  \BibitemOpen
  \bibfield  {author} {\bibinfo {author} {\bibfnamefont {T.}~\bibnamefont
  {H{\"o}rlin}},\ }\bibfield  {title} {\enquote {\bibinfo {title}
  {Deconvolution and maximum entropy in impedance spectroscopy of noninductive
  systems},}\ }\href@noop {} {\bibfield  {journal} {\bibinfo  {journal} {Solid
  State Ionics}\ }\textbf {\bibinfo {volume} {107}},\ \bibinfo {pages}
  {241--253} (\bibinfo {year} {1998})}\BibitemShut {NoStop}%
\bibitem [{\citenamefont {H{\"o}rlin}(1993)}]{horlin1993maximum}%
  \BibitemOpen
  \bibfield  {author} {\bibinfo {author} {\bibfnamefont {T.}~\bibnamefont
  {H{\"o}rlin}},\ }\bibfield  {title} {\enquote {\bibinfo {title} {Maximum
  entropy in impedance spectroscopy of non-inductive systems},}\ }\href@noop {}
  {\bibfield  {journal} {\bibinfo  {journal} {Solid State Ionics}\ }\textbf
  {\bibinfo {volume} {67}},\ \bibinfo {pages} {85--96} (\bibinfo {year}
  {1993})}\BibitemShut {NoStop}%
\bibitem [{\citenamefont {Tesler}\ \emph {et~al.}(2010)\citenamefont {Tesler},
  \citenamefont {Lewin}, \citenamefont {Baltianski},\ and\ \citenamefont
  {Tsur}}]{tesler2010analyzing}%
  \BibitemOpen
  \bibfield  {author} {\bibinfo {author} {\bibfnamefont {A.}~\bibnamefont
  {Tesler}}, \bibinfo {author} {\bibfnamefont {D.}~\bibnamefont {Lewin}},
  \bibinfo {author} {\bibfnamefont {S.}~\bibnamefont {Baltianski}},\ and\
  \bibinfo {author} {\bibfnamefont {Y.}~\bibnamefont {Tsur}},\ }\bibfield
  {title} {\enquote {\bibinfo {title} {Analyzing results of impedance
  spectroscopy using novel evolutionary programming techniques},}\ }\href@noop
  {} {\bibfield  {journal} {\bibinfo  {journal} {J. Electroceram.}\ }\textbf
  {\bibinfo {volume} {24}},\ \bibinfo {pages} {245--260} (\bibinfo {year}
  {2010})}\BibitemShut {NoStop}%
\bibitem [{\citenamefont {Boukamp}(2020)}]{boukamp2020distribution}%
  \BibitemOpen
  \bibfield  {author} {\bibinfo {author} {\bibfnamefont {B.~A.}\ \bibnamefont
  {Boukamp}},\ }\bibfield  {title} {\enquote {\bibinfo {title} {Distribution
  (function) of relaxation times, successor to complex nonlinear least squares
  analysis of electrochemical impedance spectroscopy?}}\ }\href@noop {}
  {\bibfield  {journal} {\bibinfo  {journal} {J. Phys.: Energy}\ }\textbf
  {\bibinfo {volume} {2}},\ \bibinfo {pages} {042001} (\bibinfo {year}
  {2020})}\BibitemShut {NoStop}%
\bibitem [{\citenamefont {Effendy}, \citenamefont {Song},\ and\ \citenamefont
  {Bazant}(2020)}]{effendy2020analysis}%
  \BibitemOpen
  \bibfield  {author} {\bibinfo {author} {\bibfnamefont {S.}~\bibnamefont
  {Effendy}}, \bibinfo {author} {\bibfnamefont {J.}~\bibnamefont {Song}},\ and\
  \bibinfo {author} {\bibfnamefont {M.~Z.}\ \bibnamefont {Bazant}},\ }\bibfield
   {title} {\enquote {\bibinfo {title} {Analysis, design, and generalization of
  electrochemical impedance spectroscopy (eis) inversion algorithms},}\
  }\href@noop {} {\bibfield  {journal} {\bibinfo  {journal} {J. Electrochem.
  Soc.}\ }\textbf {\bibinfo {volume} {167}},\ \bibinfo {pages} {106508}
  (\bibinfo {year} {2020})}\BibitemShut {NoStop}%
\bibitem [{\citenamefont {Liu}\ and\ \citenamefont
  {Ciucci}(2020)}]{liu2020deep}%
  \BibitemOpen
  \bibfield  {author} {\bibinfo {author} {\bibfnamefont {J.}~\bibnamefont
  {Liu}}\ and\ \bibinfo {author} {\bibfnamefont {F.}~\bibnamefont {Ciucci}},\
  }\bibfield  {title} {\enquote {\bibinfo {title} {The deep-prior distribution
  of relaxation times},}\ }\href@noop {} {\bibfield  {journal} {\bibinfo
  {journal} {J. Electrochem. Soc.}\ }\textbf {\bibinfo {volume} {167}},\
  \bibinfo {pages} {026506} (\bibinfo {year} {2020})}\BibitemShut {NoStop}%
\bibitem [{\citenamefont {Cole}\ and\ \citenamefont
  {Cole}(1941)}]{cole1941dispersion}%
  \BibitemOpen
  \bibfield  {author} {\bibinfo {author} {\bibfnamefont {K.~S.}\ \bibnamefont
  {Cole}}\ and\ \bibinfo {author} {\bibfnamefont {R.~H.}\ \bibnamefont
  {Cole}},\ }\bibfield  {title} {\enquote {\bibinfo {title} {Dispersion and
  absorption in dielectrics i. alternating current characteristics},}\
  }\href@noop {} {\bibfield  {journal} {\bibinfo  {journal} {J. Chem. Phys.}\
  }\textbf {\bibinfo {volume} {9}},\ \bibinfo {pages} {341--351} (\bibinfo
  {year} {1941})}\BibitemShut {NoStop}%
\bibitem [{\citenamefont {Davidson}\ and\ \citenamefont
  {Cole}(1951)}]{davidson1951dielectric}%
  \BibitemOpen
  \bibfield  {author} {\bibinfo {author} {\bibfnamefont {D.~W.}\ \bibnamefont
  {Davidson}}\ and\ \bibinfo {author} {\bibfnamefont {R.~H.}\ \bibnamefont
  {Cole}},\ }\bibfield  {title} {\enquote {\bibinfo {title} {Dielectric
  relaxation in glycerol, propylene glycol, and n-propanol},}\ }\href@noop {}
  {\bibfield  {journal} {\bibinfo  {journal} {J. Chem. Phys.}\ }\textbf
  {\bibinfo {volume} {19}},\ \bibinfo {pages} {1484--1490} (\bibinfo {year}
  {1951})}\BibitemShut {NoStop}%
\bibitem [{\citenamefont {Havriliak}\ and\ \citenamefont
  {Negami}(1966)}]{havriliak1966complex}%
  \BibitemOpen
  \bibfield  {author} {\bibinfo {author} {\bibfnamefont {S.}~\bibnamefont
  {Havriliak}}\ and\ \bibinfo {author} {\bibfnamefont {S.}~\bibnamefont
  {Negami}},\ }\bibfield  {title} {\enquote {\bibinfo {title} {A complex plane
  analysis of $\alpha$-dispersions in some polymer systems},}\ }in\ \href@noop
  {} {\emph {\bibinfo {booktitle} {J. Polym. Sci., Part C: Polym. Symp.}}},\
  Vol.~\bibinfo {volume} {14}\ (\bibinfo {year} {1966})\ pp.\ \bibinfo {pages}
  {99--117}\BibitemShut {NoStop}%
\bibitem [{\citenamefont {Prabhakar}(1971)}]{prabhakar1971singular}%
  \BibitemOpen
  \bibfield  {author} {\bibinfo {author} {\bibfnamefont {T.~R.}\ \bibnamefont
  {Prabhakar}},\ }\bibfield  {title} {\enquote {\bibinfo {title} {A singular
  integral equation with a generalized mittag leffler function in the
  kernel},}\ }\href@noop {} {\bibfield  {journal} {\bibinfo  {journal}
  {Yokohama Math. J.}\ }\textbf {\bibinfo {volume} {19}},\ \bibinfo {pages}
  {7--15} (\bibinfo {year} {1971})}\BibitemShut {NoStop}%
\bibitem [{\citenamefont {Garrappa}, \citenamefont {Mainardi},\ and\
  \citenamefont {Maione}(2016)}]{garrappa2016models}%
  \BibitemOpen
  \bibfield  {author} {\bibinfo {author} {\bibfnamefont {R.}~\bibnamefont
  {Garrappa}}, \bibinfo {author} {\bibfnamefont {F.}~\bibnamefont {Mainardi}},\
  and\ \bibinfo {author} {\bibfnamefont {G.}~\bibnamefont {Maione}},\
  }\bibfield  {title} {\enquote {\bibinfo {title} {Models of dielectric
  relaxation based on completely monotone functions},}\ }\href@noop {}
  {\bibfield  {journal} {\bibinfo  {journal} {Fract. Calc. Appl. Anal.}\
  }\textbf {\bibinfo {volume} {19}},\ \bibinfo {pages} {1105--1160} (\bibinfo
  {year} {2016})}\BibitemShut {NoStop}%
\bibitem [{\citenamefont {Saxena}, \citenamefont {Mathai},\ and\ \citenamefont
  {Haubold}(2004)}]{saxena2004generalized}%
  \BibitemOpen
  \bibfield  {author} {\bibinfo {author} {\bibfnamefont {R.}~\bibnamefont
  {Saxena}}, \bibinfo {author} {\bibfnamefont {A.}~\bibnamefont {Mathai}},\
  and\ \bibinfo {author} {\bibfnamefont {H.}~\bibnamefont {Haubold}},\
  }\bibfield  {title} {\enquote {\bibinfo {title} {On generalized fractional
  kinetic equations},}\ }\href@noop {} {\bibfield  {journal} {\bibinfo
  {journal} {Physica A}\ }\textbf {\bibinfo {volume} {344}},\ \bibinfo {pages}
  {657--664} (\bibinfo {year} {2004})}\BibitemShut {NoStop}%
\bibitem [{\citenamefont {Florsch}, \citenamefont {Revil},\ and\ \citenamefont
  {Camerlynck}(2014)}]{florsch2014inversion}%
  \BibitemOpen
  \bibfield  {author} {\bibinfo {author} {\bibfnamefont {N.}~\bibnamefont
  {Florsch}}, \bibinfo {author} {\bibfnamefont {A.}~\bibnamefont {Revil}},\
  and\ \bibinfo {author} {\bibfnamefont {C.}~\bibnamefont {Camerlynck}},\
  }\bibfield  {title} {\enquote {\bibinfo {title} {Inversion of generalized
  relaxation time distributions with optimized damping parameter},}\
  }\href@noop {} {\bibfield  {journal} {\bibinfo  {journal} {J. Appl.
  Geophys.}\ }\textbf {\bibinfo {volume} {109}},\ \bibinfo {pages} {119--132}
  (\bibinfo {year} {2014})}\BibitemShut {NoStop}%
\bibitem [{\citenamefont {Nigmatullin}\ and\ \citenamefont
  {Ryabov}(1997)}]{nigmatullin1997cole}%
  \BibitemOpen
  \bibfield  {author} {\bibinfo {author} {\bibfnamefont {R.}~\bibnamefont
  {Nigmatullin}}\ and\ \bibinfo {author} {\bibfnamefont {Y.~E.}\ \bibnamefont
  {Ryabov}},\ }\bibfield  {title} {\enquote {\bibinfo {title} {Cole-davidson
  dielectric relaxation as a self-similar relaxation process},}\ }\href@noop {}
  {\bibfield  {journal} {\bibinfo  {journal} {Physics of the Solid State}\
  }\textbf {\bibinfo {volume} {39}},\ \bibinfo {pages} {87--90} (\bibinfo
  {year} {1997})}\BibitemShut {NoStop}%
\bibitem [{\citenamefont {Elwakil}, \citenamefont {Allagui},\ and\
  \citenamefont {Psychalinos}(2021)}]{ieee}%
  \BibitemOpen
  \bibfield  {author} {\bibinfo {author} {\bibfnamefont {A.}~\bibnamefont
  {Elwakil}}, \bibinfo {author} {\bibfnamefont {A.}~\bibnamefont {Allagui}},\
  and\ \bibinfo {author} {\bibfnamefont {C.}~\bibnamefont {Psychalinos}},\
  }\bibfield  {title} {\enquote {\bibinfo {title} {On the equivalent impedance
  of self-similar ladder networks},}\ }\href@noop {} {\bibfield  {journal}
  {\bibinfo  {journal} {IEEE Trans. Circuits Syst. II Express Briefs}\ }\textbf
  {\bibinfo {volume} {68}},\ \bibinfo {pages} {2685--2689} (\bibinfo {year}
  {2021})}\BibitemShut {NoStop}%
\bibitem [{\citenamefont {Bateman}(1954)}]{bateman1954tables}%
  \BibitemOpen
  \bibfield  {author} {\bibinfo {author} {\bibfnamefont {H.}~\bibnamefont
  {Bateman}},\ }\href@noop {} {\emph {\bibinfo {title} {Tables of integral
  transforms}}},\ Vol.~\bibinfo {volume} {2}\ (\bibinfo  {publisher}
  {McGraw-Hill book company},\ \bibinfo {year} {1954})\BibitemShut {NoStop}%
\bibitem [{\citenamefont {Allagui}\ and\ \citenamefont {Elwakil}(2024)}]{drt}%
  \BibitemOpen
  \bibfield  {author} {\bibinfo {author} {\bibfnamefont {A.}~\bibnamefont
  {Allagui}}\ and\ \bibinfo {author} {\bibfnamefont {A.~S.}\ \bibnamefont
  {Elwakil}},\ }\bibfield  {title} {\enquote {\bibinfo {title} {Procedure for
  obtaining the analytical distribution function of relaxation times for the
  analysis of impedance spectra using the fox $h$-function},}\ }\href@noop {}
  {\bibfield  {journal} {\bibinfo  {journal} {arxiv}\ } (\bibinfo {year}
  {2024})}\BibitemShut {NoStop}%
\bibitem [{\citenamefont {Schwarz}(2005)}]{schwarz2005generalized}%
  \BibitemOpen
  \bibfield  {author} {\bibinfo {author} {\bibfnamefont {J.~H.}\ \bibnamefont
  {Schwarz}},\ }\bibfield  {title} {\enquote {\bibinfo {title} {The generalized
  stieltjes transform and its inverse},}\ }\href@noop {} {\bibfield  {journal}
  {\bibinfo  {journal} {Journal of mathematical physics}\ }\textbf {\bibinfo
  {volume} {46}} (\bibinfo {year} {2005})}\BibitemShut {NoStop}%
\bibitem [{\citenamefont {Macdonald}\ and\ \citenamefont
  {Brachman}(1956)}]{macdonald1956linear}%
  \BibitemOpen
  \bibfield  {author} {\bibinfo {author} {\bibfnamefont {J.~R.}\ \bibnamefont
  {Macdonald}}\ and\ \bibinfo {author} {\bibfnamefont {M.~K.}\ \bibnamefont
  {Brachman}},\ }\bibfield  {title} {\enquote {\bibinfo {title} {Linear-system
  integral transform relations},}\ }\href@noop {} {\bibfield  {journal}
  {\bibinfo  {journal} {Reviews of modern physics}\ }\textbf {\bibinfo {volume}
  {28}},\ \bibinfo {pages} {393} (\bibinfo {year} {1956})}\BibitemShut
  {NoStop}%
\bibitem [{\citenamefont {Fox}(1961)}]{fox1961g}%
  \BibitemOpen
  \bibfield  {author} {\bibinfo {author} {\bibfnamefont {C.}~\bibnamefont
  {Fox}},\ }\bibfield  {title} {\enquote {\bibinfo {title} {The g and h
  functions as symmetrical fourier kernels},}\ }\href@noop {} {\bibfield
  {journal} {\bibinfo  {journal} {Trans. Am. Math. Soc.}\ }\textbf {\bibinfo
  {volume} {98}},\ \bibinfo {pages} {395--429} (\bibinfo {year}
  {1961})}\BibitemShut {NoStop}%
\bibitem [{\citenamefont {Mathai}, \citenamefont {Saxena},\ and\ \citenamefont
  {Haubold}(2009)}]{mathai2009h}%
  \BibitemOpen
  \bibfield  {author} {\bibinfo {author} {\bibfnamefont {A.~M.}\ \bibnamefont
  {Mathai}}, \bibinfo {author} {\bibfnamefont {R.~K.}\ \bibnamefont {Saxena}},\
  and\ \bibinfo {author} {\bibfnamefont {H.~J.}\ \bibnamefont {Haubold}},\
  }\href@noop {} {\emph {\bibinfo {title} {The H-function: theory and
  applications}}}\ (\bibinfo  {publisher} {Springer Science \& Business
  Media},\ \bibinfo {year} {2009})\BibitemShut {NoStop}%
\bibitem [{\citenamefont {Kilbas}(2004)}]{kilbas2004h}%
  \BibitemOpen
  \bibfield  {author} {\bibinfo {author} {\bibfnamefont {A.~A.}\ \bibnamefont
  {Kilbas}},\ }\href@noop {} {\emph {\bibinfo {title} {H-transforms: Theory and
  Applications}}}\ (\bibinfo  {publisher} {CRC press},\ \bibinfo {year}
  {2004})\BibitemShut {NoStop}%
\bibitem [{\citenamefont {Mathai}\ \emph {et~al.}(1978)\citenamefont {Mathai},
  \citenamefont {Saxena}, \citenamefont {Saxena} \emph {et~al.}}]{mathai1978h}%
  \BibitemOpen
  \bibfield  {author} {\bibinfo {author} {\bibfnamefont {A.~M.}\ \bibnamefont
  {Mathai}}, \bibinfo {author} {\bibfnamefont {R.~K.}\ \bibnamefont {Saxena}},
  \bibinfo {author} {\bibfnamefont {R.~K.}\ \bibnamefont {Saxena}}, \emph
  {et~al.},\ }\href@noop {} {\emph {\bibinfo {title} {The H-function with
  applications in statistics and other disciplines}}}\ (\bibinfo  {publisher}
  {John Wiley \& Sons},\ \bibinfo {year} {1978})\BibitemShut {NoStop}%
\bibitem [{\citenamefont {Hilfer}(2002)}]{hilfer}%
  \BibitemOpen
  \bibfield  {author} {\bibinfo {author} {\bibfnamefont {R.}~\bibnamefont
  {Hilfer}},\ }\bibfield  {title} {\enquote {\bibinfo {title} {$h$-function
  representations for stretched exponential relaxation and non-debye
  susceptibilities in glassy systems},}\ }\href@noop {} {\bibfield  {journal}
  {\bibinfo  {journal} {Phys. Rev. E}\ }\textbf {\bibinfo {volume} {65}},\
  \bibinfo {pages} {061510} (\bibinfo {year} {2002})}\BibitemShut {NoStop}%
\bibitem [{\citenamefont {Gl{\"o}ckle}\ and\ \citenamefont
  {Nonnenmacher}(1993)}]{glockle1993fox}%
  \BibitemOpen
  \bibfield  {author} {\bibinfo {author} {\bibfnamefont {W.~G.}\ \bibnamefont
  {Gl{\"o}ckle}}\ and\ \bibinfo {author} {\bibfnamefont {T.~F.}\ \bibnamefont
  {Nonnenmacher}},\ }\bibfield  {title} {\enquote {\bibinfo {title} {Fox
  function representation of non-debye relaxation processes},}\ }\href@noop {}
  {\bibfield  {journal} {\bibinfo  {journal} {J. Stat. Phys.}\ }\textbf
  {\bibinfo {volume} {71}},\ \bibinfo {pages} {741--757} (\bibinfo {year}
  {1993})}\BibitemShut {NoStop}%
\bibitem [{\citenamefont {Williams}\ and\ \citenamefont
  {Watts}(1970)}]{williams1970non}%
  \BibitemOpen
  \bibfield  {author} {\bibinfo {author} {\bibfnamefont {G.}~\bibnamefont
  {Williams}}\ and\ \bibinfo {author} {\bibfnamefont {D.~C.}\ \bibnamefont
  {Watts}},\ }\bibfield  {title} {\enquote {\bibinfo {title} {Non-symmetrical
  dielectric relaxation behaviour arising from a simple empirical decay
  function},}\ }\href@noop {} {\bibfield  {journal} {\bibinfo  {journal}
  {Transactions of the Faraday society}\ }\textbf {\bibinfo {volume} {66}},\
  \bibinfo {pages} {80--85} (\bibinfo {year} {1970})}\BibitemShut {NoStop}%
\end{thebibliography}
 
%

 \end{document}